\documentclass{llncs} 
\usepackage{llncsdoc}
\usepackage{amssymb} 
\usepackage{subfig}
\usepackage{graphicx}
\usepackage[english]{babel}
\usepackage{graphicx}
\usepackage{tabularx}
\usepackage{dcolumn}
\usepackage{array}
\usepackage{pdflscape}
\begin{document} 
\title{An algorithm to simulate alternating Turing machine by signal machine}   
\author{Dawood Hasanzadeh \and Sama Goliaei}   
\institute{University of Tehran, Tehran, Iran.}
\maketitle
\begin{center}
{d.hasanzadeh67, sgoliaei} @ut.ac.ir
\end{center}
\begin{abstract} 
Abstract Geometrical Computation as a new model of computation is the counterpart of Cellular Automata that has Turing computing ability. In this paper we provide an algorithm to simulate Alternating Turing Machine in the context of Signal Machine using techniques adopted from the features of Signal Machine to set up and manage the copies/branches of Alternating Turing Machine. We show that our algorithm can simulate Alternating Turing Machine in Signal Machine as same functionality as classic family of Turing Machines. Time complexity of the algorithm is linear as ordinary simulated Turing Machines. Depending on the computation tree space complexity is exponential order of \textit{d}, where \textit{d} is the depth of the computation tree.
\end{abstract} 
\markboth{An Algorithm to Simulate Alternating Turing Machine in Signal Machine}{An Algorithm to Simulate Alternating Turing Machine in Signal Machine} 
\section{Introduction}
\textit{Automata} are machines that repetitively execute pre-determined instructions. \textit{Cellular Automata} (CA) is a discrete model of computation consisting of a set of regular cells, each in one of the finite possible states such as \textit{on} or \textit{off}. The set of cells around each cell called its \textit{Neighborhood} which determines the state of the cell.  At the beginning ($ t=0 $) an initial state is assigned to each cell, after each time step a new generation of cells produces based on pre-determined \textit{Rules}. These rules determine the new states of cells according to their current states and neighborhood's states of each cell. The rules are fixed and apply to the whole grid simultaneously. The \textit{space-time diagram} of CA is the integer numbers for the space and the natural numbers for the time.

Using Euclidean geometry for computation forms a new model of computation known as \textit{Abstract Geometrical Computation} (AGC) which first introduced by Jerome Durand-Lose in 2003 \cite{durand2003calculer}. AGC is an analog model of computation and the counterpart of  (CA). As Durand-Lose mentions AGC does not just come "out of the blue" because of its CA origins \cite{durand2005abstract}. If we replace the integer and natural numbers by real and positive real numbers respectively, we gain the continuous model of AGC where cells/particles are dimensionless resulting the continuous time and space. In this structure dimensionless particles move forward/backward in the continuous space-time diagram with constant speeds. Movement of particles is along real numbers for the space and only upward for the time, indeed time is always upward and there is no way to move back on the time axes.

Consider the moving particles as Euclidean lines called \textit{signals}; each signal is a sample of a \textit{meta-signal} with constant speed where the number of meta-signals is finite. In fact a signal is a piece of information spreads on the space-time diagram for a certain purpose. A cross point forms when two (or more) signals collide/meet each other, which is called a \textit{collision}. When signals collide, \textit{collision rules} determine aftermath of that collision, e.g., what new signals should be produced instead of previous signals before the collision. If for the collision of some signals there is not a predefined collision rule, it called a \textit{blank} collision and signals continue their path without any changes in direction and type. The described structure is called \textit{Signal Machine} (SM) in the context of AGC. SM is composed of three main parts: meta-signals, their speeds and collision rules. 

Since its introduction in 2003, some computations and simulations has been done in this model. It is proved turing-computing ability could be carried out through two-counter automata in SM model \cite{durand2005abstract}. Conservative abstract geometrical computation \cite{durand2004abstract,durand2006abstract} is a model that can simulate any turing machine and decide any recursively enumerable problem by creating accumulation points. NP-complete problems such as SAT problem can be solved efficiently in SM \cite{duchier2010fractal}. For this purpose in a division process the space slices to shape a comb to solve the SAT problem. The time and space are bounded in this structure. In a massively parallel manner it is possible to solve Q-SAT problem too \cite{duchier2010massively}. In \cite{duchier2012computing} the writers proposed a particular generic machine to solve Q-SAT using Map/Reduce paradigm. As the proposed machine is modular it is possible to solve satisfiability variants such as SAT, $ \texttt{\#} $SAT and MAX-SAT.

The simulation of ordinary TMs is presented in \cite{durand2004abstract} for classical computation; in this article Durand-Lose presents a model in SM to decide semi-decidable problems according to Black Hole model of computation. The size of a TM is an important issue that has been considered so far \cite{baiocchi2001three,kudlek1996small}. In the context of SM this is addressed in \cite{durand2011abstract}, small signal machines, able to perform fully classical computation (TM) with regards to the number of meta-signals and collision rules are presented. Other types of TM like type-2 Turing machine (T2-TM) is presented in a mixed representation of real numbers plus an exact value in (-1, 1) \cite{durand2011abstractB} and reversible TM is simulated in \cite{durand2012abstract}. If a construction be rational accumulation points coordinates, time and space, are computably enumerable numbers (c.e. numbers) and difference of two such numbers (d-c.e. numbers) respectively \cite{durand2012abstractB}.  Accumulation points as a limit of a sequence of signals, with regard to the number of different present speeds is discussed in \cite{becker2013abstract}.

Alternating Turing Machine (ATM) is a generalization of Turing Machine which in a specified state may choose more than one on-going state and each state labelled by a \textit{universal} or \textit{existential} quantifier. In this article we focus on ATM in the context of AGC. Hence, an algorithm is introduced to simulate ATM in SM model directly. The idea is to set up a scaffold to start copy of the branches of ATM: when the copy initialization ends the computation starts to freezing to the right and left, then the pre-set unfreeze signal restarts the computation at both sides. Finally, when all the computations end, final results send to the \textit{result collector} signal that is the answer of ATM. 

Section 2 provides backgrounds and the formal definitions of SM, configuration, space-time diagram, SM techniques, TM and ATM. The copy initialization process, unfreezing, recovery,  scaling of the computation, the process of collecting the results and time and space complexities are provided in section 3. Article ends by conclusion and future works  in section 4.

\section{Backgrounds}
According to the continuous feature of signal machines they can compute and simulate CA computations and even more complex computations that CA can not afford like semi-decidable problems. SMs are kind of \textit{collision based computing} where particle colliders are signals, collision rules run the collisions and therefore guide the computations.
\subsection{Definitions}
\begin{definition} (Signal Machine) A \textit{Signal Machine (SM)} is composed of three parts \textit{(M, S, R)} where \textit{M} defines a finite set of \textit{meta-signals}, \textit{S} is a mapping from \textit{M} to $\Bbb{R}$  which defines the \textit{speed} of each meta-signal, and, \textit{R} is a function from the subset of \textit{M} (at least two meta-signals) into a subset of \textit{M} which defines the \textit{collision rules} of signals.
\end{definition}

Signals are instances of meta-signals and there may be many signals in an initial configuration but the number of signals is finite. Signals have constant speeds and the collision rules determine what happens when two or more signals collide. A collision happens by at least two signals. Each rule is defined as a pair of (input meta-signals, output meta-signals). If there is not a rule for a collision, signals continue their way exactly as before of the collision. Time and space are continuous and the moving particles are dimensionless.
 
\begin{definition} (Configuration) A \textit{Configuration (c)} is a function 
 maps $ \Bbb{Q} $ (for rational machines) and $ \mathbb{R} $  (for real machines) to 
$ M \cup R $ (Meta-signals and Rules) such that the set of all possible particles is finite i.e., all signals and collisions are isolated.
\end{definition}
Definition 2 illustrates that there is a maximum countable number of signals and collisions. In fact we have a infinitive space that just a finite part of it is used to form a configuration. If we assume each mapping of the mentioned function is a line,  we are talking about a one dimensional space.

\begin{definition} (Space-Time Diagram) A space-time diagram is made of the exact choices that  repeatedly performed by the machine. 
In each time step, a new configuration is added to the last one and the space time diagram evolves repeatedly. Time is always upward and the space is either positive or negative \cite{durand2012abstractB}.
\end{definition}

\begin{figure}
\centering
\includegraphics[scale=0.3]{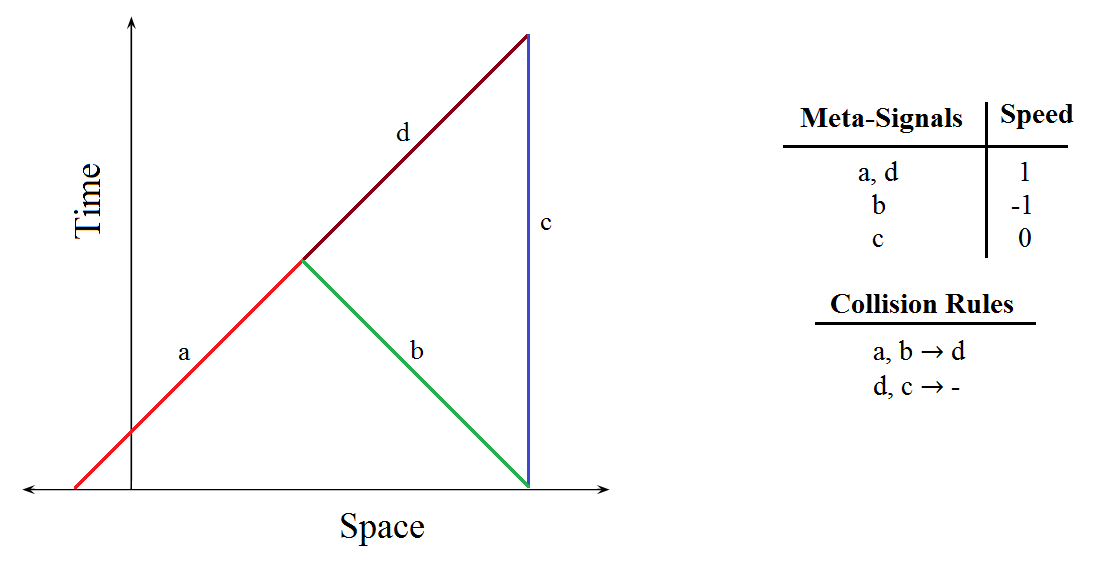}
\caption{Space-time diagram with a simple example}
\label{fig:spacetime}
\end{figure}
Figure \ref{fig:spacetime} shows a simple example of space-time diagram on the left, meta-signals and collision rules on the right. As we see the time scale is $ \mathbb{R}^+ $ and the space scale is $ \mathbb{R} $. Signal $ \texttt{a} $ starts in a negative real number position. There is four meta-signals and one sample of each one on the space-time diagram; signals $ \texttt{a, d} $ of speed 1, signal $ \texttt{b} $ of speed -1 and signal $ \texttt{c} $ of speed 0. when $ \texttt{a} $ collides $ \texttt{b} $, according to the first collision rule signal $ \texttt{d} $ produces until it collides $ \texttt{c} $, in this location by collision of $ \texttt{d} $ and $ \texttt{c} $ according to the second collision rule nothing produces, i.e., the collision is a \textit{void} collision.
\subsection{Turing Machine and Signal Machine} \label{TM}
Turing Machine (TM) is one of the abstract classical models of computation which temporary stores the inputs on a tape. Addition to the tape, TM is defined by a set of states, a finite set of symbols, transition function and the head.  The tape is made of cells, each cell is capable of holding just one symbol of the defined alphabet. The read-write head traverses the tape and in every step reads the cell symbol, according to it's transition function writes the exchange and goes to the left or right.

Officially, a TM is defined by TM =
($Q$, $q_{i}$, $ \Gamma $,   $^ \wedge $, $ \texttt{\#} $, $ \delta $) 
where Q is a finite set of states, $ q_{i} $ is the initial state,
 $ \Gamma $ is a finite set of symbols, $ ^\wedge $ 
 is the head, 
 $ \texttt{\#} $
 is blank symbol and
$ \delta: Q \times \Gamma \rightarrow Q \times \Gamma \times \{\leftarrow, \rightarrow \} $
is the transition function.  Here we are not going to have a detailed study on TMs, the reader is referred to \cite{linz2011introduction}.

As showed in \cite{durand2005abstract} the model of Signal Machine has the power of Turing-Computation. The Turing-Computation power of signal machine is proved by simulating any two-counter automaton. Therefore signal machine is a model of computation has at least Turing-Computing capability.

The simulation of different types of TM is presented in \cite{durand2011abstract,durand2011abstractB,durand2012abstractB,durand2009abstractK}. The simulation of ordinary TM is as follows: some signals of speed zero encode the symbols on the tape and a signal of non-zero speed guides the left-right movements of the head, defining the state of machine \cite{durand2011abstract}.

\begin{figure}[ht]
\centering
\includegraphics[scale=0.44]{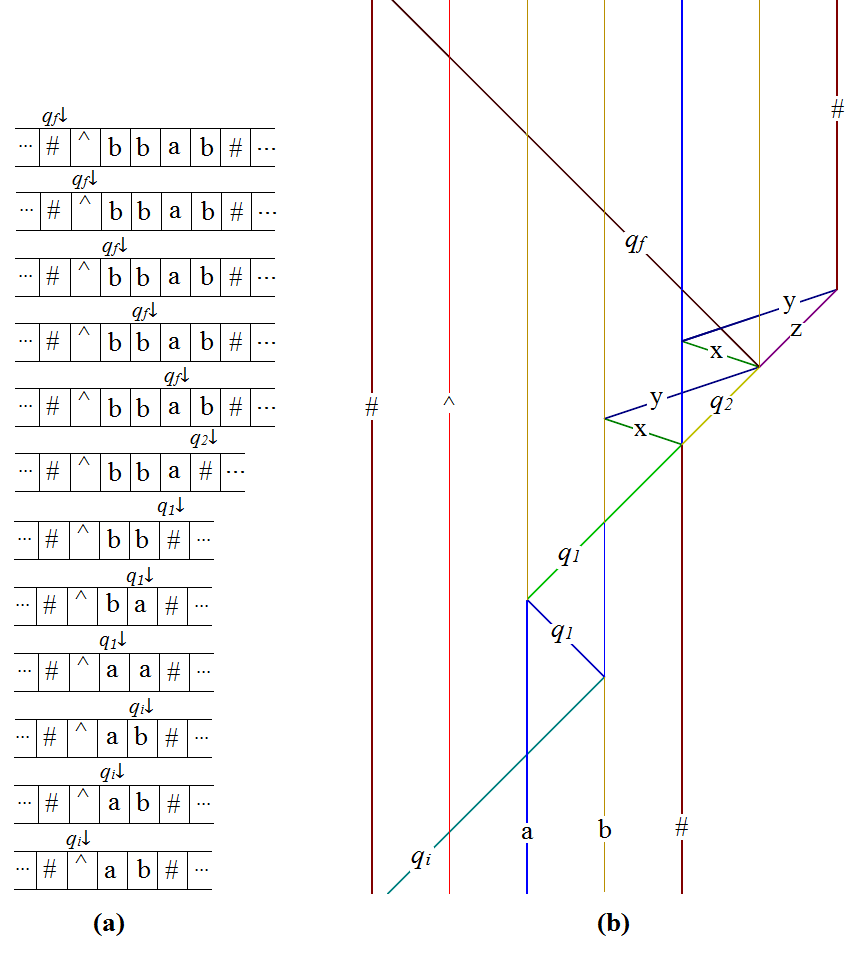}
\caption{Ordinary TM computation (a) TM transitions (b) Equivalent SM }
\label{fig:tm}
\end{figure}

Figure \ref{fig:tm} shows the simulation of TM in SM. Left shows the transitions of TM on the tape and right shows the SM equivalent simulation. Signal
$ \texttt{q}_{i} $ is the initial state of TM. One can guess the collision rules easily from Figure \ref{fig:tm}. Signals \texttt{x} and \texttt{y} provide the possibility of tape extension. Head signals (\texttt{q}) have speed 1 (left to right) and -1 (right to left). Signal $ \texttt{\#} $ shows the blank symbol or border of computation. The final result (final state of TM), $ \texttt{q}_{f} $, diffuses to the left of computation.

\subsection{Signal Machine Techniques}
In this part we introduce some useful techniques of SM introduced before in the literature. The idea of our algorithm is based on these techniques and we use them to set up a structure to simulate ATM in SM. These techniques include computing the middle of any computation, freezing and unfreezing of a computation and scaling of computation. However we will change and modify these techniques to gain our objectives later.
\subsubsection*{Middle of Computation}
\begin{proposition} (Middle of Computation) Middle of a computation can be easily computed by propagating three signals, two with the same absolute values of speeds (3x and -3x) and the third with 1/3 speed of the other two, i.e., 1x.
\end{proposition}
\begin{figure}[ht]
\centering
\includegraphics[scale=0.33]{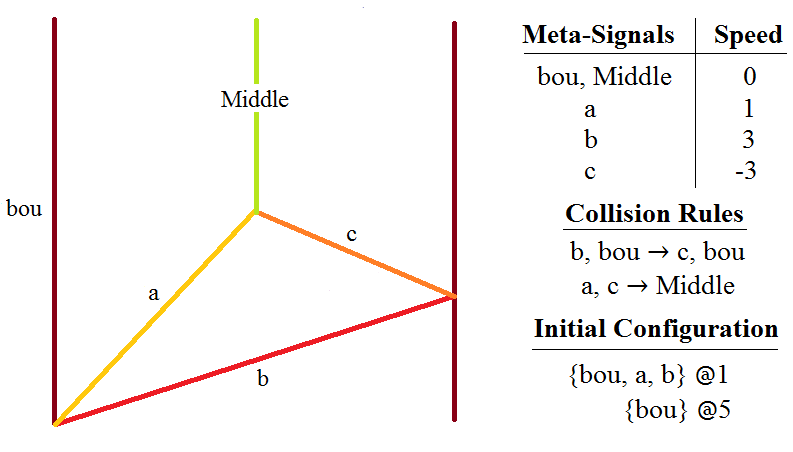}
\caption{Computing the middle}
\label{fig:middle}
\end{figure}
The computation of the middle is some times necessary, e.g,. in \cite{duchier2012computing} the middle indefinitely computed to shape a scaffold for solving SAT problem. 
Figure \ref{fig:middle} shows the details of computing the middle of any computation. Meta signals, speeds, collision rules and initial configuration are listed on the right. At the beginning, four signals shape the initial configuration then according to the collision rules the middle of computation will be marked by $ \texttt{Middle} $ signal. Obviously, location of $ \texttt{Middle} $ is $ x = 3 $, exactly the middle of computation.

\subsubsection*{Freezing and Unfreezing}
\begin{proposition} (Freezing/Unfreezing) By adding a new meta-signal, \texttt{f}, with a speed greater than any present one it is possible to redirect the computation signals so that by occurring a collision between \texttt{f} and any computation signal, say \texttt{c},  \texttt{f} is reproduced and \texttt{c} is replaced by a constant speed signal. Unfreezing is the opposite \cite{durand2009abstractK}.
\end{proposition}

\begin{figure}[ht]
\centering
\includegraphics[scale=0.55]{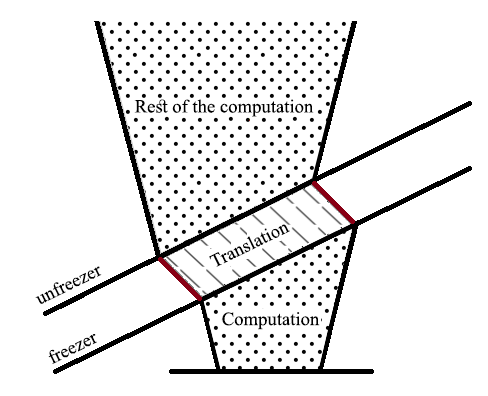}
\caption{Freezing/Unfreezing operation}
\label{fig:freezing}
\end{figure}

Figure \ref{fig:freezing} illustrates the freezing/unfreezing operation. The speed of \texttt{freezer} is greater than all computation signals and the speed of the frozen signals (the area between the parallel dark red lines) is less than \texttt{freezer} speed. Depending on the speed of \texttt{freezer} it is possible to conduct the computation to the left (positive speed) or right (negative speed). As collisions are just simply points, they will be frozen and unfrozen by \texttt{freezer} and \texttt{unfreezer} signals respectively. The computation goes ahead normally above the \texttt{unfreezer}. The freezing and freezer signals have to be parallel to accurately recall the computation after freezing. Translation part shows the frozen parallel signals (Figure \ref{fig:freezing}), colliding to the unfreezer the computation restarts, in fact the translation area is a delay to stop the computation for a special purpose like conduct the computation signals.

\subsubsection*{Scaling the Computation}
\begin{proposition} (Scaling the Computation) A frozen computation can be redirected and therefore scaled according to this idea that the unfreezing signal has a smaller speed \cite{durand2009abstractK}. 
\end{proposition}
\begin{figure}[ht]
\centering
\includegraphics[scale=0.55]{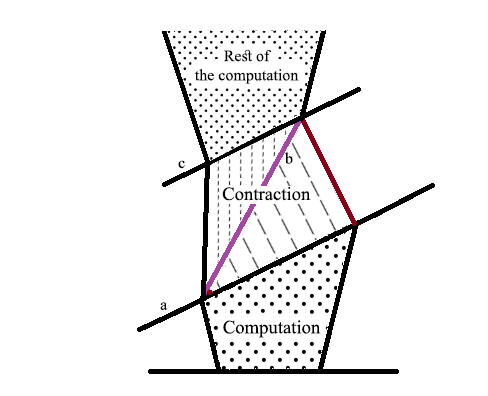}
\caption{Scaling operation}
\label{fig:scaling}
\end{figure}
Figure \ref{fig:scaling} shows scaling operation of any computation: up-going computation is redirected twice to gain a scaled computation. Signal $ \texttt{a} $ freezes the computation by directing it to the left for the first time, signal $ \texttt{b} $ redirects the computation and signal $ \texttt{c} $ recovers the scaled computation.

In the contraction area the computation signals will be frozen and ready to be contracted to gain an optionally scaled computation depending on speed of signal \texttt{b}. Signal \texttt{c} unfreezes the contracted signals as usual and the rest of the computation, scaled down, goes ahead above signal \texttt{c}.

After these preliminaries, in the coming sections we introduce the TMs, ATMs and related simulations. 

\subsection{Alternating Turing Machine}
Alternating Turing Machine (ATM) is a Non-Deterministic TM, introduced by Chandra and Stockmeyer \cite{chandra1976alternation} and Kozen \cite{kozen1976parallelism}  in 1976. In 1981 in an article named Alternation \cite{hopcroft1981alternation} they, together, made a comprehensive introduction to ATM.

An ATM  is defined exactly as TM but with two extra special states, denoted by $ "\exists " $  and $ "\forall " $. In state $ \exists $ machine accepts the input if and only if at least one of the successors accepts. In state $ \forall $ the machine accepts the input if and only if all of the successors accept.

\begin{figure}[ht]
\centering
\includegraphics[scale=0.33]{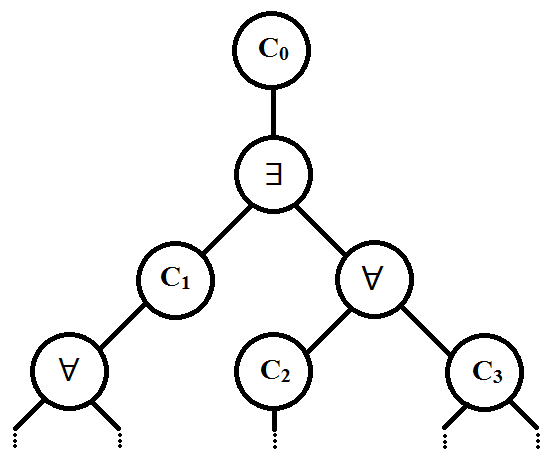}
\caption{Computation tree of an ATM}
\label{fig:atm}
\end{figure}

Figure \ref{fig:atm} shows a schematic example of an ATM.  In the configuration tree each node is labeled with a possible configuration of the machine. Each edge is a transition from machine's current configuration to the reachable ones. Since the final result is given when the root's computation finishes, and the root's computation depends on it's descendants and so on, the final result is determined recursively. ATM accepts the computation on a specified input if and only if the root of the computation tree is accepted.

It is obvious a non-deterministic turing machine is an ATM with only $\exists$ states. In a configuration with only one child the node accepts if and only if its child accepts.  In Figure \ref{fig:atm}, $C_0$ accepts if and only if one of the five paths to the terminal nodes (leaves) is acceptable.

\begin{figure}[ht]
\centering
\includegraphics[scale=0.33]{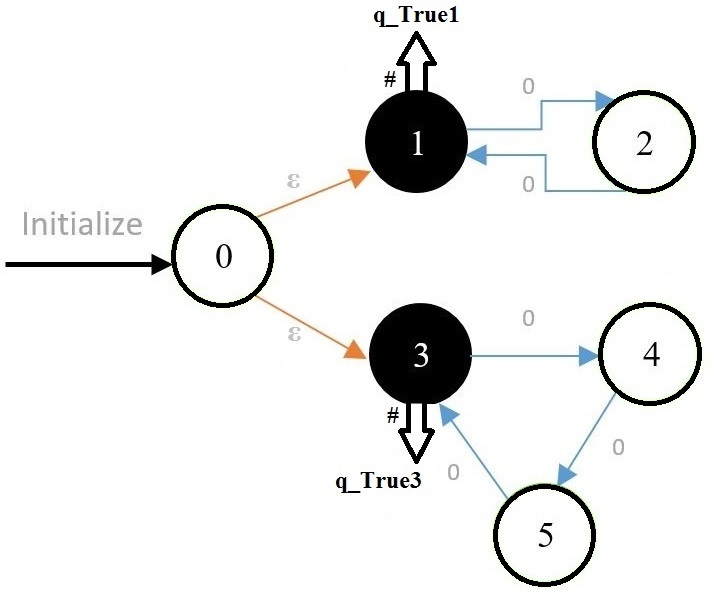}
\caption{State transition of ATM for two and three divisible number of zeros}
\label{fig:trs}
\end{figure}

Before addressing the algorithm, we need to know the states and transition table of our ATM. Figure \ref{fig:trs} shows the transition state of an ATM that decides a boolean string if having zeros divisible by two and three. The numbers inside the circles indicate the names of states; 1 for $ q_1 $ and so on. $ q_1 $ and $ q_3 $ are the final states: if ATM stops on both of these two states the final answer is $ \textsc{Yes} $ and the number of zeros of the string is divisible by two and three. For example strings 1011000100 and 000000 are acceptable and strings 0110011 and 0000000 are not acceptable by this machine.

\begin{table*}
\centering
\caption{Transition table of ATM with two final states $q_{true1}$ and  $q_{true3}$}
\label{transitiontable}
\begin{tabular}{>{\centering}p{1.4 cm}>{\centering}p{1.4 cm}>{\centering}p{2 cm}>{\centering}p{2 cm}>{\centering}p{2 cm}>{\centering}p{2 cm}}
\hline
$\Gamma$: \{ $\exists$, $\forall$ \} & $\delta$ &  $\widehat{•}$  & $0$ & $1$ & $\texttt{\#}$    \tabularnewline \hline
$\forall$ & $q_0$  &  $q_1$, $q_3$,$\widehat{ }$, $\rightarrow$ & -- &  --  &  --  \tabularnewline
$--$ & $q_1$ & $--$ & $q_2$, $0$, $\rightarrow$ & $q_1$, $1$, $\rightarrow$ &$ q_\mathrm{true1}$, $\texttt{\#}$, $\leftarrow$ 	\tabularnewline
$-- $& $q_2$ & $--$ & $q_1$, $0$, $\rightarrow$ & $q_2$, $1$, $\rightarrow$ & $q_\mathrm{false1}$, $\texttt{\#}$, $\leftarrow$  \tabularnewline
$-- $& $q_3$ &$ --$ & $q_4$, $0$,$ \rightarrow$ &$ q_3$,$ 1$,$ \rightarrow$ & $q_\mathrm{true3}$,$ \texttt{\#}$, $\leftarrow$  \tabularnewline
$-- $& $q_4$ &$ --$ &$ q_5$,$ 0$, $\rightarrow$ & $q_4$, $1$, $\rightarrow$ & $q_\mathrm{false3}$,$ \texttt{\#}$, $\leftarrow$  \tabularnewline
$-- $& $q_5$ &$ --$ & $q_3$,$ 0$, $\rightarrow$ & $q_5$,$ 1$, $\rightarrow$ & $q_\mathrm{false3}$, $\texttt{\#}$, $\leftarrow$  \tabularnewline

$--$ & $q_\mathrm{true1}$ & $ q_\mathrm{true1}$ ,$\widehat{ }$, $\leftarrow$ & $q_\mathrm{true1}$, $0$, $\leftarrow$ &$ q_\mathrm{true1}$, $1$, $\leftarrow$ & $--$  \tabularnewline
$--$ & $q_\mathrm{true3}$ & $q_\mathrm{true3}$, $\widehat{ }$, $\leftarrow$ & $q_\mathrm{true3}$, $0$, $\leftarrow$ & $q_\mathrm{true3}$, $1$, $\leftarrow$ & $--$  \tabularnewline
$-- $& $q_\mathrm{false1}$ & $q_\mathrm{false1}$, $\widehat{ }$, $\leftarrow$ & $q_\mathrm{false1}$, $0$, $\leftarrow$ & $q_\mathrm{false1}$, $1$, $\leftarrow$ & $--$  \tabularnewline
$--$ & $q_\mathrm{false3}$ & $q_\mathrm{false3}$, $\widehat{ }$, $\leftarrow$ & $q_\mathrm{false3}$, $0$, $\leftarrow $& $q_\mathrm{false3}$, $1$, $\leftarrow$ & $--$  \tabularnewline
\hline
\end{tabular}
\end{table*} 
As Figure \ref{fig:trs} shows the upper and the lower branches compute divisibility by two and three respectively. Machine starts in an initial configuration, at first because zero is divisible by two and three, both branches are in final states $ q_1 $ and $ q_3 $. By reading each cells content if it was zero machine changes it's state. Notice that the quantifier of $ q_0 $ where the computation branches is universal, means both branches must return $ \textsc{Yes} $ value for a string to be accepted. 

Table \ref{transitiontable} shows the transition function of Figure \ref{fig:trs}.  When the machine encounters the $ \texttt{\#} $ sign at the end of the string, decides it, if it was in states $ q_1 $  or $ q_3 $  returns $ q_\mathrm{true1} $  and $ q_\mathrm{true3} $  respectively, otherwise it returns $ q_\mathrm{false1} $ and $ q_\mathrm{false3} $ respectively.

Now we are ready to represent the simulation of ATM in SM. To do this, as in ordinary TM we start normally until the machine needs to branch or copy itself. The algorithm is given step by step in the next coming sections.
\section{Simulation of ATM by SM}
The main issue of simulation is to manage the copies or branches of ATM. At first we handle the copy initialization to reach the pre-defined on-going states by a presented structure. After that when whole computation freezes to the desired directions, unfreezing operation starts by adding two special unfreezer signals added to the right and left of the computation (proposition 2). The freezing/unfreezing operation, here, is different because of the intrinsic complexity of the structure. Unfreezing makes the branches restart their computations. Final answer of ATM is given by a signal called \textit{result collector} where retrieves the results of each computation.

Coming subsections present the mentioned steps in details; subsections 3.1 - 3.5 respectively provide the copy initialization, computation recovery, collecting the results, the whole simulation and finally the complexity of the algorithm.

\subsection{Computation Copy Initializing}
The middle of computation can be simply computed (Proposition 1). Computing the middle helps to freeze/unfreeze the computation accurately. Suppose the machine needs to be copied when the head of machine is on a cell in the middle of computation; as we will see later for orderly freeze/unfreeze signals it is necessary to start from a corner, hence the middle helps to do this accurately.

\begin{figure}[ht]
\centering
\includegraphics[scale=0.47]{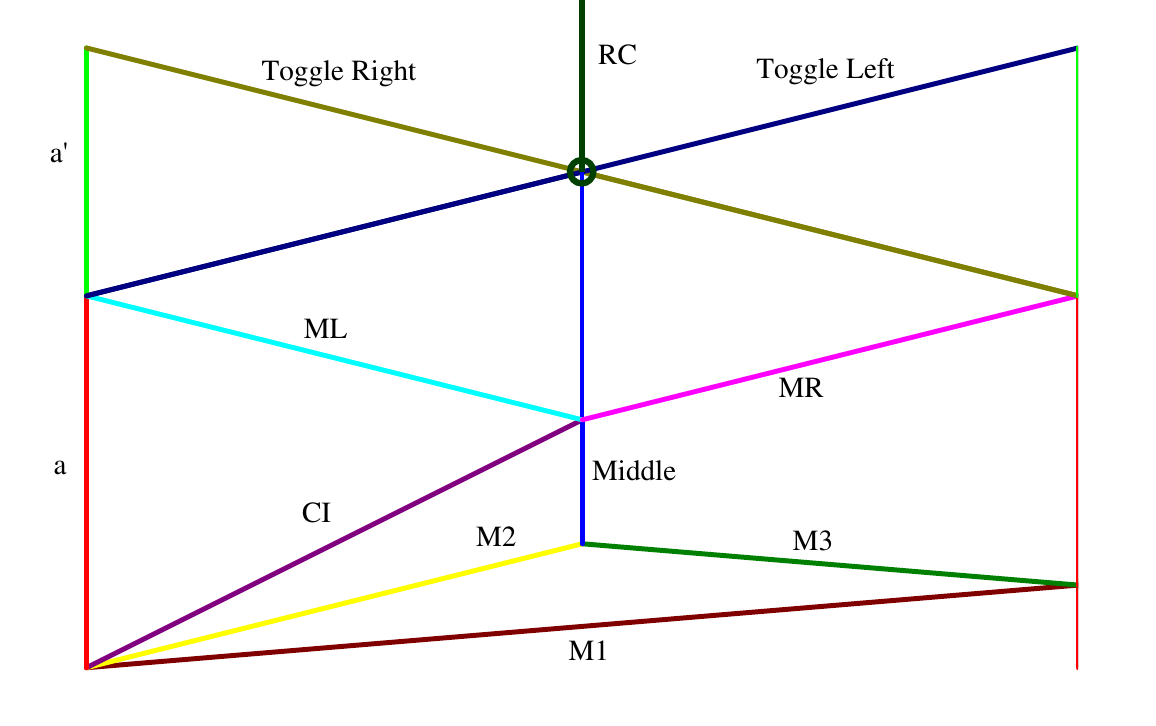}
\caption{Structure of copy}
\label{fig:copyinit}
\end{figure}

There should be a structure to initialize the copy operation when it is time to copy the configuration, i.e., machine is in a state that needs to branch. Figure \ref{fig:copyinit} provides such a structure. The structure is different from an ordinary copy (in fact freeze) structure.

The copy structure, as in Figure \ref{fig:copyinit}, the first action is to compute the middle; this is done by sending $ \texttt{M1} $ and $ \texttt{M2} $  signals from the left. Notice that these two signals as they may be needed in any moment must have the highest speed between the rest of signals. Signal \texttt{a} from both sides is the boundary of computation.

Suppose at the beginning copy operation should be initialized e.g., the second row of Table 1 for $ q_0 $. Signal $ \texttt{CI} $ is sent as Copy Initiator; when it collides the $ \texttt{Middle} $, $ \texttt{MR} $ and $ \texttt{ML} $ are sent with opposite equal speeds to the right and left of the computation respectively. $ \texttt{MR} $ and $ \texttt{ML} $ reach to the boundaries simultaneously as the amount of their speed is equal (but opposite) and  they sent from the middle of computation (see Figure \ref{fig:copyinit}).

When $ \texttt{MR} $ and $ \texttt{ML} $ collide the boundaries, $ \texttt{Toggle Left} $ and $ \texttt{Toggle Right} $ are generated again with opposite equal speeds. These signals are supposed to freeze the computation to the left and right respectively. Tha reason of intentionally conversion of signal $ \texttt{Middle} $ to $ \texttt{RC} $, meaning Result Collector, will be noticed soon.   

$ \texttt{Toggle} $ signals act as a mirror to reflect coming signals to the desired directions. Let us divide the configuration into three parts: the $ Middle $ part (signal $ \texttt{Middle} $), the $ right $ part which is from signal $ \texttt{Middle} $ to the right boundary (signal $ \texttt{a} $) and the $ left $  part which is from signal $ \texttt{Middle} $ to the left boundary (signal $ \texttt{a} $). These three parts should freeze to both left and right sides. The important tip here is every signal has to be frozen two times, so when a signal meets the first $ \texttt{Toggle} $, freezes to the left (or right) and it has to be continued to be frozen by the second $ \texttt{Toggle} $. This is true for $ right $ and $ left $ parts.

\begin{figure}[ht]
\centering
\includegraphics[scale=0.33]{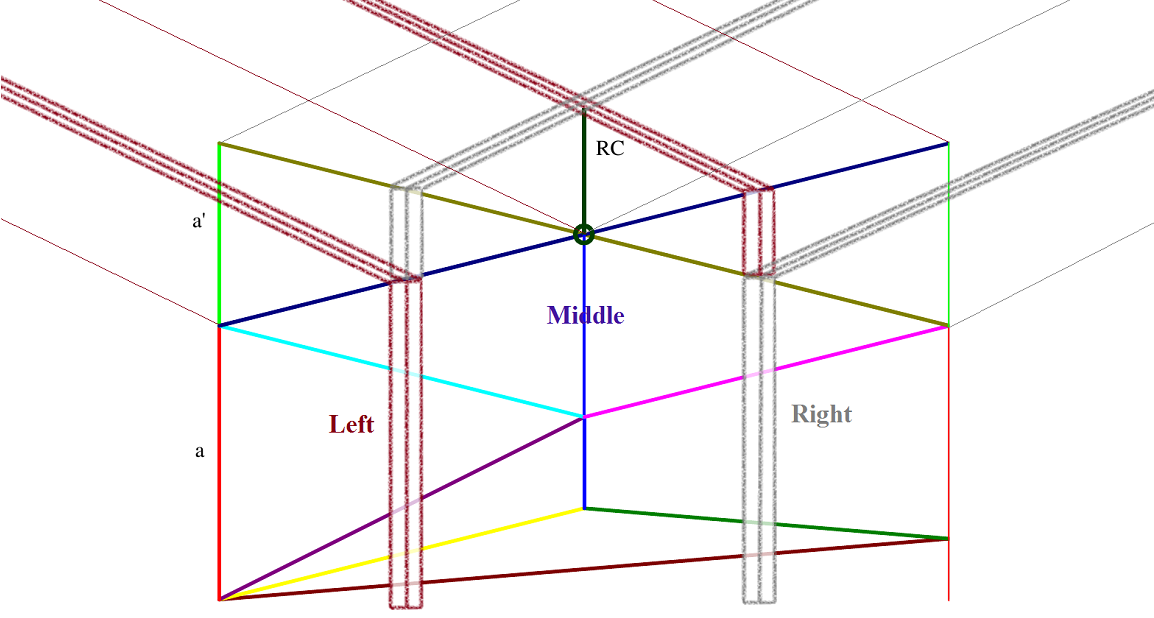}
\caption{Freezing the configuration to the left and right}
\label{fig:freeze2side}
\end{figure}

Figure \ref{fig:freeze2side} shows the described freezing operation. The $ right $ and $ left $ possible signals showed by Gray and dark Red respectively. Focus on the $ right $ part of configuration; when signals collide the first $ \texttt{Toggle} $ which is $ \texttt{Toggle Right} $,  the color of signals changes to be continued (copied) and the signals freeze to the right. When the continued signals collide the second $ \texttt{Toggle} $ which is $ \texttt{Toggle Left} $, they only freeze to the left. Freezing operation of the $ left $ part is the same as the $ right $ part but by opposite speeds.

Notice that signals $ \texttt{Toggle Right} $ and $ \texttt{Toggle left} $ reach the $ Middle $ at the same time and $ \texttt{Middle} $ directly freezes to both sides when it collides them both; in fact $ Middle $ has no need to be copied. About the boundary signal $ \texttt{a} $, freezes and continues at the same time with generation of the first toggle in both sides.

\subsection{Unfreezing, Computation Recovery and Scaling}
\begin{figure}[ht]
\centering
\includegraphics[scale=0.33]{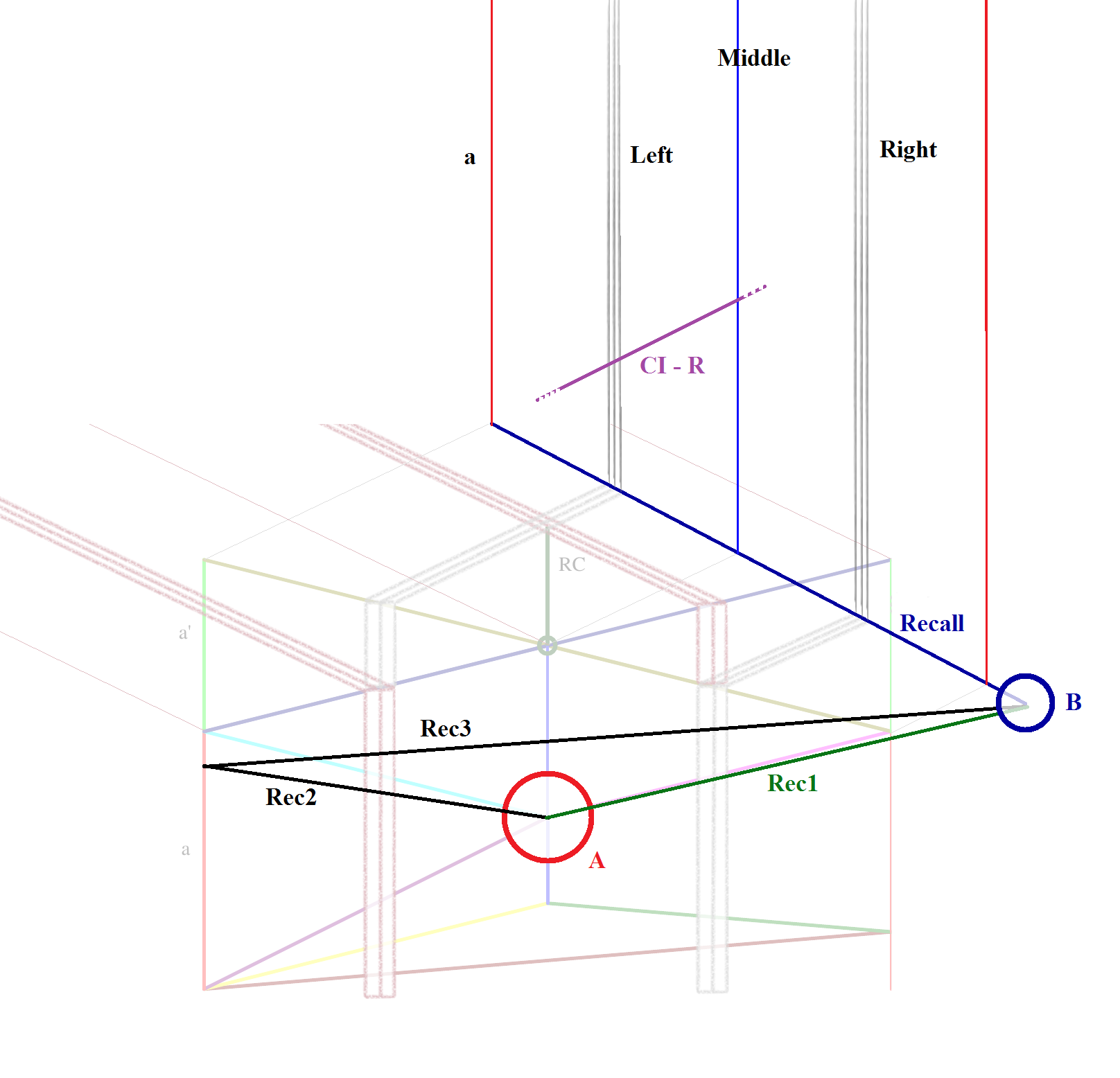}
\caption{The process of recovery and scale for right frozen configuration}
\label{fig:unfreezeandscale}
\end{figure}

After freezing configuration to the both sides it is time to recover it. For this purpose we need four signals, three for locating a point to send the forth signal and recover the configuration. The first three signals must have a speed more than any configuration signal, to be able to collect all frozen signals and no signal left because of low speed. Figure \ref{fig:unfreezeandscale} shows such a structure.

According to Figure \ref{fig:unfreezeandscale} signals $ \texttt{Rec1} $, $ \texttt{Rec2} $, $ \texttt{Rec3} $ and final recovering signal named $ \texttt{Recall} $ are used to recover the configuration. As mentioned before, if point $ B $, that is the recovering start point, be higher than the first frozen signal $ \texttt{a} $ because of low speed, then $ \texttt{a} $ is not recoverable. Signal $ \texttt{Racall} $ starts to unfreeze any frozen signal encounters from point $ B $ and by colliding the second boundary, $ \texttt{a} $, disappears.

Signals $ \texttt{Rec1} $ and $ \texttt{Rec2} $ are sent from point $ A $, that is exactly the same point that $ \texttt{MR} $ and $ \texttt{ML} $ were sent to generate $ \texttt{Toggle} $ signals. In fact the recovery operation has been started at the same time that $ \texttt{Toggle} $'s were going to generate. Thus we can be sure the frozen signals are recoverable. It is worth mentioning that the recovery operation does not affect/conflict the freezing operation; the reason is that the speed of recovering signals is a little more than the freezing related ones.

Scaling the configuration (shrink or even stretch) is done when we recover the frozen signals by $ \texttt{Recall} $. The more speed of $ \texttt{Recall} $, the higher is scaling and the narrower is new configuration, i.e., the configuration shrinks more. In Figure \ref{fig:unfreezeandscale} right configuration scaled by about 3/4 of origin configuration, as well as for the left configuration.

Taken together, when $ \texttt{CI} $ collides $ \texttt{Middle} $, seven signals generate: one for continuing $ \texttt{Middle} $ as $ \texttt{RC} $, two for generating $ \texttt{Toggle Left} $ and $ \texttt{Toggle Right} $, two for finding location of point $ B $ to generate $ \texttt{Recall} $ and two for generating left counterpart of $ B $. Signal $ \texttt{CI-R} $ is the continuation of computations were in the origin configuration, now transferred to the right to compute one of the branches. It has a counterpart, $ \texttt{CI-L} $, will be mentioned in the next section.
\subsection{Collecting the Results}
In general the purpose of copying a configuration is to branch the computations and at the end collect the results of the branches to decide a certain problem. For this, two important questions should be taken into account: the first is that how to restart the computations? and the second is that how to collect the results of new configurations and retrieve the final answer of machine? In this section we discuss these two questions and offer the final operation giving the final result.

Signals $ \texttt{CI-R} $ and $ \texttt{CI-L} $ are representative of one of the states that machine wanted to go (Table \ref{transitiontable}). For example as it showed in Table \ref{transitiontable} when the machine arrives to sign $\;\, \widehat{•}\;\, $, wants to transit to states $ q_1 $ and $ q_3 $; therefore signals $ \texttt{CI-R} $ and $ \texttt{CI-L}$ can be continuation of $ q_1 $ or $ q_3 $. Thus computation goes on with $ \texttt{CI-R} $ in the right computation and $ \texttt{CI-L}$ in the left computation. We at first know the copy initiator state (here $ q_0 $), thus when it is time to recover computations, in each branch we recover one of the desired states (here $ q_1 $ or $ q_3 $).

\begin{figure}[ht]
\centering
\includegraphics[scale=0.3]{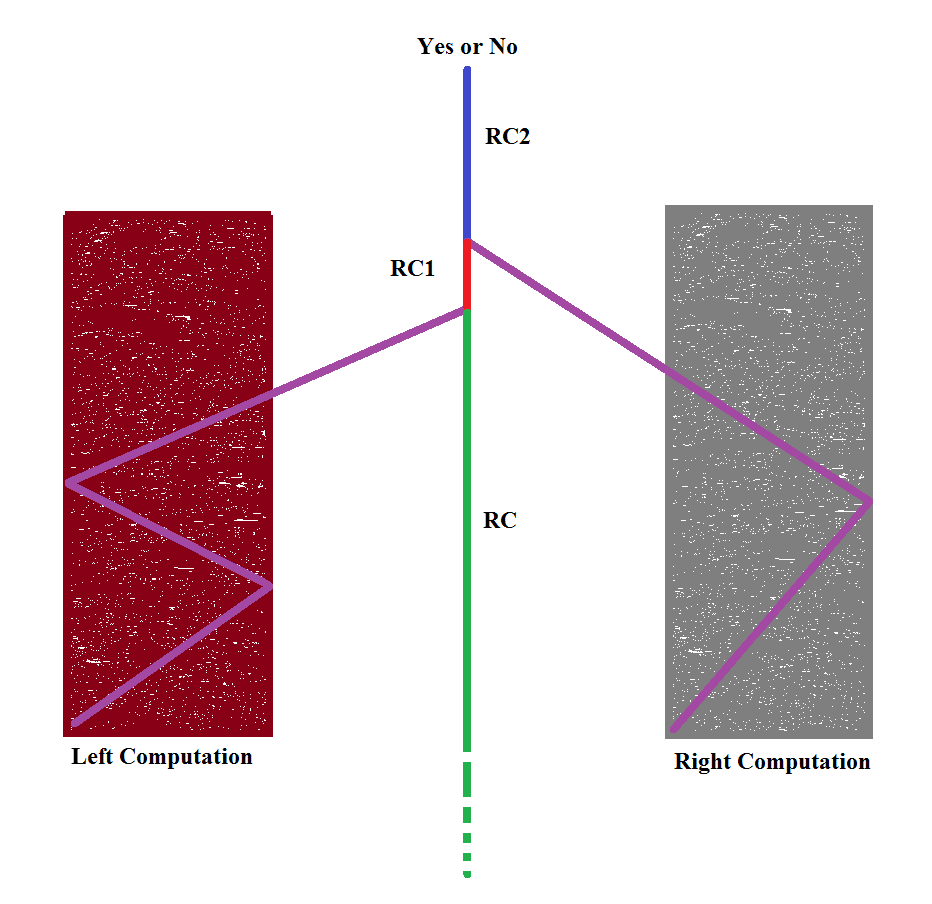}
\caption{Collecting final results}
\label{fig:resultcollect}
\end{figure}

Now we suppose all computations have reached their final states and have an answer to offer. To illustrate this consider the results, shown in Figure \ref{fig:resultcollect} which shows two computations gained their final results; the right side shows computation of the machine that restarted computing by $ q_1 $ and the left side shows the same with $ q_3 $. Signal $ \texttt{RC} $ is the signal which generated instead of $ \texttt{Middle} $. The collision between final result of left computation and $ \texttt{RC} $ generates $ \texttt{RC1} $ to show that one of results retrieved, then $ \texttt{RC1} $ collides the final result of right computation and $ \texttt{RC2} $ generates as the final result of all computations which is the answer of machine for an input string. If machine accepts the string the answer is $ \textsc{Yes} $ and if rejects the string the answer is $ \textsc{No} $.

Note that Figure \ref{fig:resultcollect} shows only one step of collecting operation. Left and right computations may have sub-branches that will be hierarchically collected to offer the final answer.

\subsection{ATM and SM}
In this section we discuss the aforementioned scaffolds on an ATM. Figure \ref{fig:trs} and Table \ref{transitiontable} are the state transition and transition table of an ATM respectively. As mentioned in section \ref{TM} this ATM decides a boolean string having the number of zeros divisible by both two and three.

\begin{figure}[ht]
\centering
\includegraphics[scale=0.5]{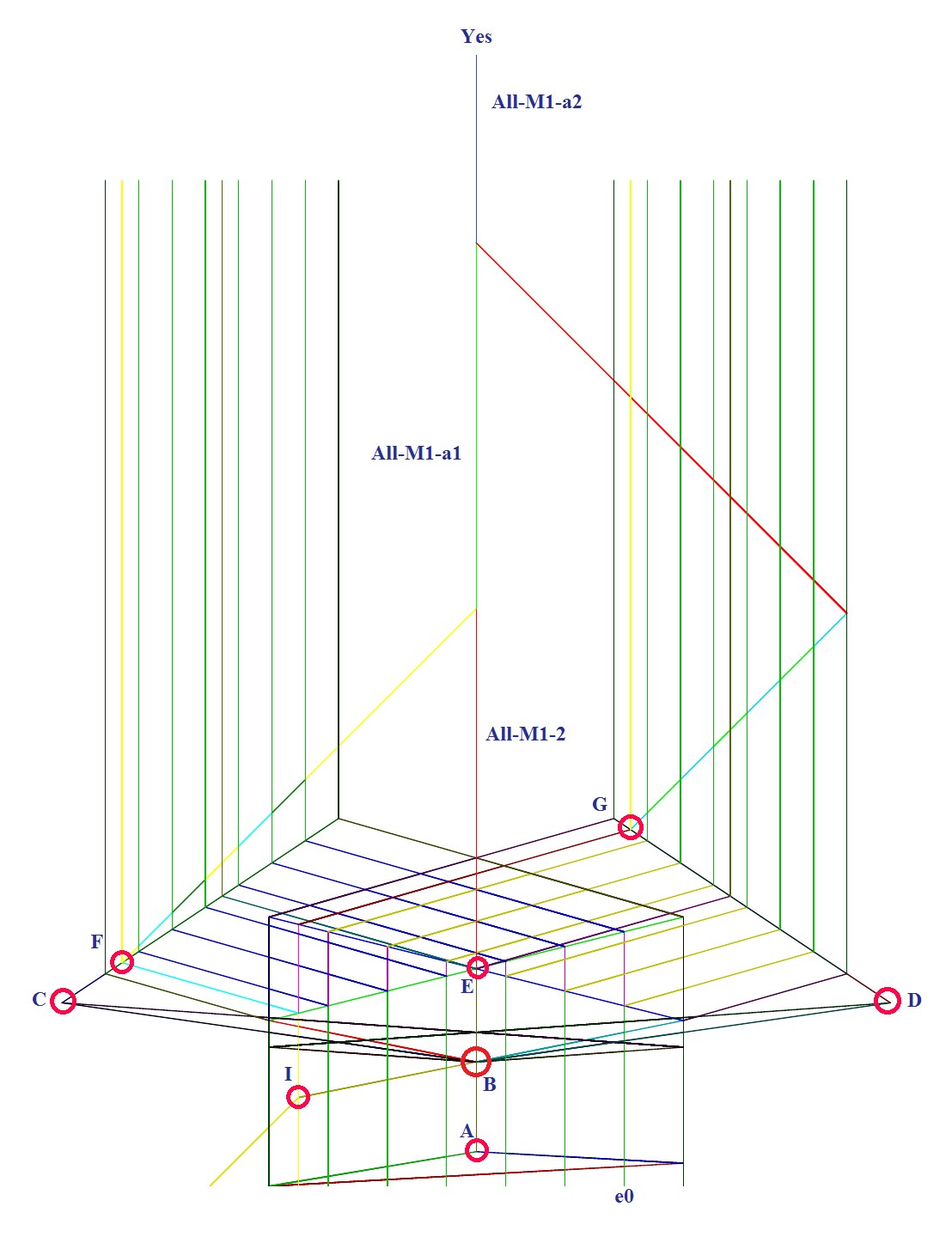}
\caption{Corresponding SM of the ATM with input 000000 and final answer $ \textsc{Yes} $}
\label{fig:ATMinSM}
\end{figure}

Figure \ref{fig:ATMinSM} shows the simulating of an ATM in SM that decides input 000000 having the number of zeros divisible by two and three; the final answer of this computation is $ \textsc{Yes} $. For convenient eight points in Figure \ref{fig:ATMinSM} highlighted to show the simulation steps. At first we should find the middle of computation; this is done in point $ A $. The point $ I $ is where a signal send to collide $ \texttt{Middle} $ and start the copy operation; after this collision in point $ B $, seven signals generate. The points $ C $ and $ D $ generated to restrain frozen signals by sending a recovering signal. In points $ F $ and $ G $ the location of the head, i.e., the possible states are recovered.

Point $ E $ is a strategic point for collecting the final results, when we used $ \texttt{Middle} $ to freeze the configuration then there is no need for it to be continued. But we need an extra signal to collect the coming results, so we replace $ \texttt{Middle} $ with a special signal, $ \texttt{All-M1-2} $, which shows the origin configuration ($ M1 $) has a universal quantifier, waiting for it's answers.

We use a standard to determine this kind of signal's names for transparency. This standard has three parts: first part determines the quantifier type, second part defines which sub-machine demanded for copy and third part defines the number of copies. Therefore $ \texttt{All-M1-2} $ reveals sub-machine $ M1 $ with a universal quantifier is waiting for it's two pending results.

When the illustrated machine in Figure \ref{fig:ATMinSM} retrieves first result from the left, the collector signal $ \texttt{All-M1-2} $ transits to $ \texttt{All-M1-a1} $ to signal retrieving the first result, at last by retrieving the last result from right transits to $ \texttt{All-M1-a2} $ and returns the final answer $ \textsc{Yes} $ for the specified string 000000.
\subsection{Complexity}
Space-time diagram can be viewed as a network consisting of a \textit{Directed Acyclic Graph} (DAG) with \textit{nodes} (collisions) and \textit{links} (signals). Therefore time complexity is defined as longest path or maximal number of sequence of collisions in the network and called \textit{collisions depth}. Accordingly space complexity is the number of pairwise un-related signals \cite{duchier2012computing}.

For ordinary TM the space complexity is $ 2 |Q| + |\Gamma| + 5 $: two meta-signal for every state that goes to the left and right of the computation, $ \Gamma $ signals for encoding the tape symbols and five signals to handle the tape enlargement i.e. signals \texttt{x, y, z, z'} and \texttt{\#}. \texttt{z'} is counterpart of \texttt{z} that is not shown in Figure \ref{fig:tm}. Therefore the space complexity is $S_{TM} = \mathcal{O}(Q + \Gamma) $. Because one collision happens in each time step regardless of the enlargement collisions, time complexity of TM is linear i.e. $T_{TM} = \mathcal{O}(n) $. 

Considering Figure \ref{fig:atm}, the tree of computation which each node is a computation on its own could branch to at most $ 2^{d-1} $ where \textit{d} is the depth of computation tree. Space complexity of ATM is $S_{ATM} = \mathcal{O}(2^d . (Q + \Gamma)) = 2^d \times S_{TM} $. Time complexity of ATM, maximal sequential collisions, is linear as same as ordinary TM:  $T_{ATM} =  T_{TM} $.  Depth of computation, \textit{d}, depends on number of branches which depends on special states. If there be \textit{m} special states like $ q_{0} $ in Table 1, each of which branch to \textit{k} branches, then d = m + 1 and the computation tree has $ k^{d-1} $ leaves should be considered in the complexities.
\section{Conclusion and future works}
We show how to simulate ATM in the context of AGC using characteristics of this geometrical model of computation (SM). Manipulation of the SM techniques allows making a structure to perform the proceedings of ATM. The simulation goes by applying modified techniques to shape the desired structure. All simulation steps is presented as an algorithm.

The time and space complexity of our algorithm is calculated according to the ordinary TM. As all branches perform the computations simultaneously, regardless a bit of delay of copy operation, the time complexity of ATM is the same as ordinary TM. Space complexity of simulated ATM is depending on, \textit{d}, the depth of computation tree e.g., it is exponential order of \textit{d}. 

Finding a better equivalent SM to an ATM is an effort we do not know about it yet. There may be better SMs for simulating ATMs. Two dimensional (2D) simulating as in CA and SM has its own limitations compelling the simulation to the more complex situation. Three dimensional (3D) simulations may provide faster results and it is worth of consideration. Therefore testing and simulating of many problems including types of TMs may be considered in future researches. 
 \bibliography{ref.bib}
 
\end{document}